\documentclass[twocolumn,showpacs,amsmath,amssymb]{revtex4}
\usepackage{graphicx}
\begin{document}
\title{Polymers with attractive interactions on the Husimi lattice}
\author{Pablo Serra}
\email{serra@famaf.unc.edu.ar}
\homepage{http://tero.fis.uncor.edu/~serra}
\affiliation{Facultad de Matem\'atica, Astronom\'{\i}a y F\'{\i}sica\\
Universidad Nacional de C\'ordoba\\
Ciudad Universitaria - 5000 C\'ordoba\\
Argentina}
\author{J\"urgen F. Stilck}
\email{jstilck@if.uff.br}
\affiliation{Instituto de F\'{\i}sica\\
Universidade Federal Fluminense\\
Av. Litor\^anea s/n\\
24210-340 - Niter\'oi, RJ\\
Brazil}
\author{Welchy L. Cavalcanti}
\author{Kleber D. Machado}
\email{kleber@fisica.ufsc.br}
\affiliation{Departamento de F\'{\i}sica\\
Universidade Federal de Santa Catarina\\
88.040-900 - Florian\'opolis - SC\\
Brazil} 
\date{\today}

\begin{abstract}
We obtain the solution of models of self-avoiding walks with
attractive interactions on Husimi lattices built with squares. Two
attractive interactions are considered: between monomers on
first-neighbor sites and not consecutive along a walk and between
bonds located on opposite edges of elementary squares. For
coordination numbers $q>4$, two phases, one polymerized the other
non-polymerized, are present in the phase diagram. For small values
of the attractive interaction the transition between those phases is
continuous, but for higher values a first-order transition is found.
Both regimes are separated by a tricritical point. For q=4 a richer
phase diagram is found, with an additional (dense) polymerized phase,
which is stable for for sufficiently strong interactions between bonds. The
phase diagram of the model in the three-dimensional parameter space displays
surfaces of continuous and discontinuous phase transitions and lines of
tricritical points, critical endpoints and triple points.
\end{abstract}

\maketitle

\section{Introduction and definition of the model} 
\label{intro}
Self-avoiding walks have been found to be useful models for the study of the 
behavior of polymers for quite a long time \cite{dg79}. The
self-avoidance constraint in general makes these models difficult to
solve on regular lattices. On lattices with hierarchical treelike
structure, however, like the Bethe \cite{b82} and the Husimi
\cite{h50} lattices, it is not difficult to solve such models
exactly. From the point of view of critical phenomena, these
solutions lead to classical or ideal chain critical exponents, but
non-universal features of the phase diagram may be closer to the ones
observed on regular lattices than those provided by the usual
mean-field methods \cite{g95}.

The effect of monomer-solvent interactions on the behavior of
polymers diluted in poor solvents may be included in the model by
allowing attractive interactions between segments of the chains. This
induces  a competition between repulsive, excluded volume
interactions and the attractive short range interactions. As these
latter interactions become sufficiently strong, the chains may change
from an extended to a collapsed state \cite{dg75}. While this
collapse transition, usually identified with the $\Theta$-point,
appears as a tricritical point in mean-field approximations and
non-classical approximations on three-dimensional lattices, in
two dimensions the situation does not seem to be so simple.
Transfer-matrix calculations \cite{bn89} and exact Bethe-ansatz
results \cite{bnw89} for a $O(n)$ model with four-spin interactions on
the square lattice lead two phase diagrams where the second order
transition line between the polymerized and non-polymerized phases
ends at a multicritical point whose precise nature is not clear from
these calculations, but which is definitely not a tricritical
point. In the limit $n \rightarrow 0$ this model corresponds to
self-avoiding walks with attractive interactions between bonds of the
walk which are located on opposite edges of elementary squares of the
lattice. The four-spin interactions in the magnetic model are related
to interactions between bonds on the corresponding polymer model. On
the other hand, studies of the behavior of self-avoiding walks on the
square lattice with attractive interactions between monomers located
on first-neighbor sites but not consecutive along the walk point to a
tricritical collapse transition \cite{ds85}. The main
motivation of this work was to address the question of apparently
quite similar models of attractive walks (bonds or monomers
interacting) leading to qualitatively distinct phase diagrams. The problem of
models with interactions between monomers or bonds only on Husimi lattices was
considered in the  literature \cite{sms96,p02,ssm02}. On a four coordinated
($q=4$) Husimi tree, the phase 
diagram  of the model with monomer-monomer interactions is qualitatively
similar to the one found in general when $q>4$. In the parameter space defined
by the activity of a monomer ($x$) and the Boltzmann factor of the elementary
interaction between bonds ($\omega$), a non-polymerized phase is stable at low
values of $x$, whereas a polymerized phase is found at higher activities. The
transition between those phases is continuous at low values of $\omega$, but
becomes discontinuous as $\omega$ is increased. These two regimes are separated
by a tricritical point. When the interactions are between bonds (Boltzmann
factor $\kappa$), a third phase is stable in part of the parameter space. This
phase is a dense phase (all sites are visited by the polymer) and
the transition continuous transition line between the non-polymerized and
the polymerized phases ends at a critical endpoint. The phase transition
between the polymerized and the dense phases may be continuous or not, a
tricritical point being found on this transition line. Transfer matrix
calculations for this model on the square lattice suggest that this picture
may be observed there as well \cite{mos01}.

We solve a model of interacting self-avoiding walks on Husimi
lattices (core of the Husimi trees), built with squares \cite{h50}.
At each site of the lattice the ramification of squares is equal to
$\sigma$, and therefore the coordination number will be
$q=2(\sigma+1)$. Solutions of models on such lattices may be
considered approximations to the solution on hipercubic lattices of
the same coordination number, so that a Husimi lattice with
$\sigma=1$ leads to a solution which approximates the one on a square
lattice, the solution for $\sigma=2$ may be an approximation for the
solution on the cubic lattice, and so on. The Husimi lattice solution
may be considered to be the third member of a sequence of
approximations whose first two are regular mean-field and Bethe
lattice solutions.  In a mean-field calculation no correlations are
taken into account, while Bethe lattice and Husimi lattice solutions
short range correlations are taken care of. It should be stressed,
however, that all these solutions lead to classical critical
exponents. In some cases, it has been shown that Bethe and Husimi
lattice solutions may show features of the phase diagram of models on
regular lattices which are not present in the corresponding
mean-field approximations \cite{g95,ws92,bs95,p02b}.

\begin{figure}
\includegraphics[scale=0.7]{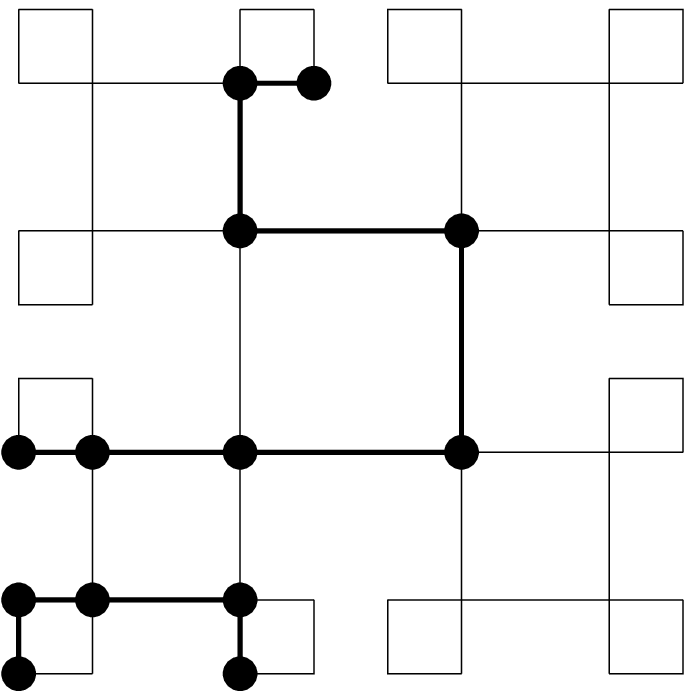}
\caption{A Husimi tree with $\sigma=1$ and three generations with
two polymers on it. The statistical weight of this configuration is
equal to $x^{11}\omega^3\kappa^2$.}
\label{f1}
\end{figure}

The model we consider is of self-avoiding walks on a Husimi tree,
with the initial and final monomer of each walk located at the
surface of the tree, so that the density of endpoint monomers in the
core of the tree is always equal to zero. We associate an activity
$x$ to each bond of a walk, and include an interaction energy
$\epsilon_m$ for each pair of monomers on first neighbor sites of the
lattice with no bond of the walk between them. Also, an interaction
energy $\epsilon_b$ is associated to each pair of bonds located on
opposite edges of elementary squares. The grand partition function of
the model on a lattice with $N$ sites may be written as
\begin{equation}
Y(x,\omega,\kappa;N)=\sum x^{N_b}\omega^{N_{im}}\kappa^{N_{ib}},
\label{e1}
\end{equation}
where the sum is over all configurations of the walks on the lattice,
$N_b$ is the number of bonds in the configuration, $N_{im}$ is the
number of pairs of interacting monomers, and $N_{ib}$ is the number
of interacting bonds. The Boltzmann factors which correspond to the
interactions are given by $\omega=\exp(-\epsilon_m/k_BT)$ and
$\kappa= \exp(-\epsilon_b/k_BT)$. A configuration of walks and the
corresponding statistical weight may be found in Fig. \ref{f1}. 

\section{Solution of the model on the Husimi tree}
\label{solution}
\begin{figure}
\includegraphics[scale=0.8]{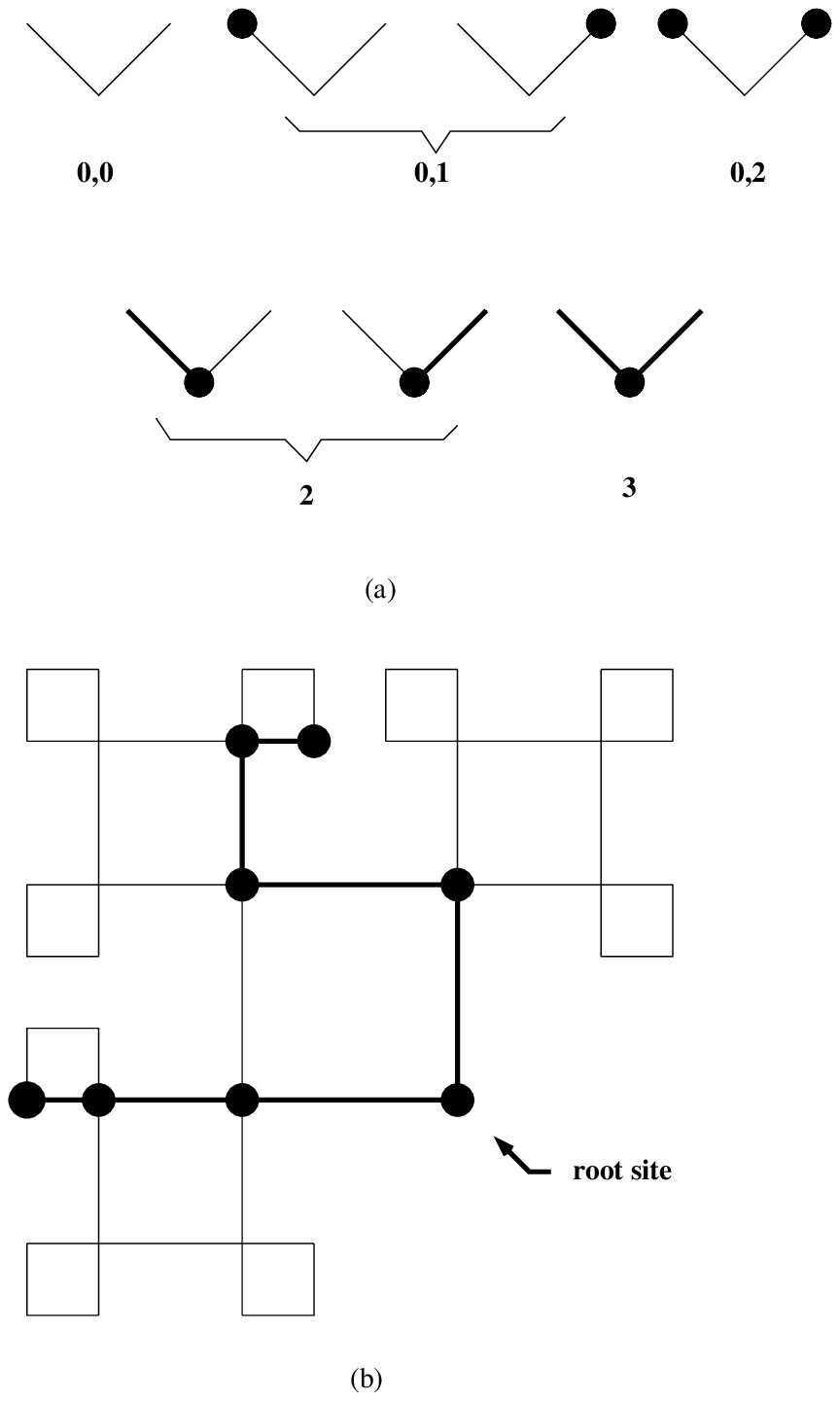}
\caption{(a) The possible configurations of the root site of a
subtree. (b) A 3-generations subtree is built attaching three
2-generations subtrees to the vertices of the new root square. In
this example $\sigma=1$. Two 2-generation subtrees have root
configuration 1, while the other one has root configuration 0,0. The
resulting root configuration of the 3-generations subtree is 2.}
\label{f2}
\end{figure}

To solve the model on the Husimi lattice we use a recursive
procedure, defining subtrees of the Husimi tree and establishing
recursion relations between the partial partition functions of the
model on the subtrees, for fixed configurations of the root site.
Fig. \ref{f2} shows the three possible root configurations of a
subtree, labeled by the number of bonds incident at the root site
from above, and a diagram illustrating how to obtain the partial
partition functions of a $(n+1)$-generations subtree from the partial
partition functions of $n$-generations subtrees. Initially, we consider three
subcases for the root configuration with no incident bond, defined by the
number of monomers in the first neighbor sites to the root site (0, 1, or
2). To obtain the recursion relations for $g_{0,0}$, $g_{0,1}$, $g_{0,2}$,
$g_1$, and $g_2$, we consider all the possibilities of attaching
three sets of $\sigma$ $n$-generations subtrees to the vertices of
the elementary square at the root of the new $(n+1)$-generations
subtree. The recursion relations are obtained so that the activity of
the bonds and the Boltzmann factors of the interactions between
monomers and bonds in the elementary square at the root of the new
$(n+1)$-generations subtree are considered in the iteration. We then notice
that the partial partition functions for the configuration with no incident
bond appear in only two combinations in the recursion relations, which are:
\begin{equation}
g_0=g_{0,0}+g_{0,1}+g_{0,2}
\end{equation}
and
\begin{equation}
g_3=g_{0,0}+\omega g_{0,1}+\omega^2 g_{0,2}.
\end{equation}
If we call
$g_i$, $i=0,1,2,3$ the partial partition functions of the model
defined on an $n$-generations subtree and $g_i^{\prime}$ the same functions on
a $n+1$-generations subtree, we may write the recursion
relations as
\begin{subequations}
\label{rr}
\begin{eqnarray}
g_0^{\prime}&=&g_0^{3\sigma}+3 Hg_0^{2\sigma}+
(1+2\omega)H^2g_0{\sigma} + \nonumber \\
 &&2xF^2 g_0^{\sigma}+\omega^2H^3+2x\omega F^2 H+x^2 F^2 g_3^{\sigma},
\label{e4}\\
g_1^{\prime}&=&2xF(g_0^{2\sigma}+2\omega H g_0^{\sigma}+\omega^3H^2+
xg_0^{\sigma}g_3^{\sigma}+ \nonumber \\
&&x\omega^2Hg_3^{\sigma}+x^2\omega\kappa
 g_3^{2\sigma} 
+x\omega^2\kappa F^2), \label{e5}\\
g_2^{\prime}&=&x^2F^2(g_0^{\sigma}+\omega^2H+2x\omega\kappa g_3^{\sigma}),
\label{e6}\\
g_3^{\prime}&=&g_0^{3\sigma}+3 Hg_0^{2\sigma}+ 2 
\omega(Hg_0^{2\sigma}+ \nonumber \\
&&\omega H^2 g_0^{\sigma}+xF^2 g_0^{\sigma})+
\omega^2(H^2 g_0^{\sigma}+\omega^2H^3+ \nonumber \\
&&2x\omega F^2H+x^2F^2g_3^{\sigma};
\label{e7}
\end{eqnarray}
\end{subequations}
where
\begin{equation}
F=\sigma g_1 g_3^{(\sigma-1)}, \label{e7n}
\end{equation}
and
\begin{equation}
H=\sigma g_2 g_3^{(\sigma-1)} + \frac{\sigma(\sigma-1)}{2}
g_1^2 g_3^{(\sigma-2)}. \label{e8}
\end{equation}

It is convenient to define the ratios
\begin{equation}
a=\frac{g_1}{g_0},\; b=\frac{g_2}{g_0}, \;\mbox{and} \; c=\frac{g_3}{g_0},
\label{rat}
\end{equation}
which obey the following recursion relations
\begin{subequations}
\label{rrr}
\begin{eqnarray}
a^{\prime}&=&2xf(1+2\omega h+\omega^3 h^2+x c^\sigma+ \nonumber \\
&&x \omega^2 c^\sigma h+x^2\omega\kappa c^{2\sigma}+
x\omega^2\kappa f^2)/q, \label{rra}\\
b^{\prime}&=&x^2f^2(1+\omega^2h+2x\omega\kappa c^\sigma)/q,\label{rrb}\\
c^{\prime}&=&(1+h+2\omega h+3\omega^2h^2+2\omega x f^2+\nonumber \\
&&\omega^4h^3+2x\omega^3f^2h+x^2\omega^2c^\sigma f^2)/q, \label{rrc}
\end{eqnarray}
\end{subequations}
where
\begin{subequations}
\label{para}
\begin{eqnarray}
f&=&\sigma a c^{(\sigma-1)}, \label{ef}\\
h&=&\sigma b c^{(\sigma-1)}+\frac{\sigma(\sigma-1)}{2}
a^2 c^{(\sigma-2)},\; \mbox{and}\label{eh}\\
q&=&1+3h+(1+2\omega)h^2+2xf^2+ \nonumber \\
 & &\omega^2h^3+2x\omega f^2h+x^2c^\sigma f^2.\label{eq}
\end{eqnarray}
\end{subequations}

\begin{figure*}
\includegraphics{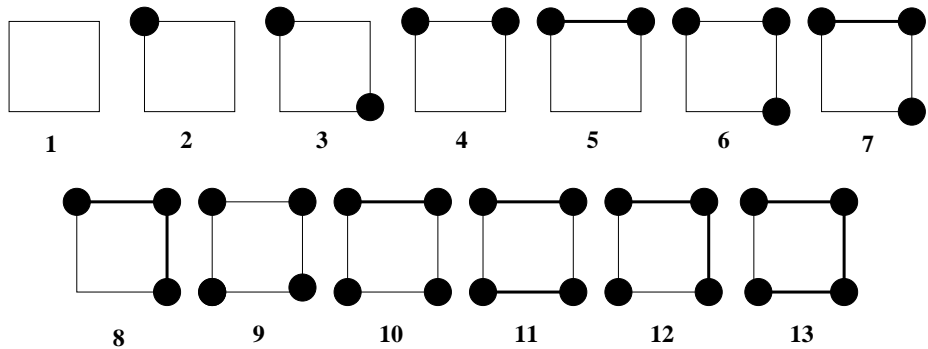}
\caption{Possible configurations of the central square of the tree.
Each of them contributes with a monomial in the calculation of the
partition function of the model on the Husimi tree.}
\label{f3}
\end{figure*}

The partition function of the model on the Husimi tree may then be
obtained if we consider the operation of attaching four sets of
$\sigma$ subtrees to the central square of the tree.
In Fig. \ref{f3} the contributing configurations of the central square
are shown in the order of the corresponding monomials appearing in
the resulting expression below
\begin{eqnarray}
Y_n(x,\omega,\kappa)&=&g_0^{4\sigma}+4g_0^{3\sigma}H+
2g_0^{2\sigma}H^2+4\omega g_0^{2\sigma}H^2+ \nonumber \\
&&4xg_0^{2\sigma} F^2+
4\omega^2g_0^\sigma H^3+ \nonumber \\
&&8x\omega g_0^\sigma F^2H+
4x^2g_0^\sigma g_3^\sigma F^2+\omega^4H^4+ \nonumber \\
&&4x\omega^3F^2H^2+
2\kappa x^2\omega^2F^4+ \nonumber \\
&&4x^2\omega^2g_3^\sigma F^2H+
4x^3\omega\kappa g_3^{2\sigma}F^2.
\label{pf}
\end{eqnarray}
We expect the thermodynamic behavior of the model on the Husimi tree
to be quite different from the one found on regular lattices, since
the surface sites dominate in the thermodynamic limit, when the 
number of iterations 
$n \rightarrow \infty$ \cite{mhz74}. We therefore will focus our 
attention on the
behavior in the central region of the tree, which we will refer to as
the Husimi lattice\cite{h50}. Considering the contributions to the
partition function in Eq. \ref{pf}, we may calculate the mean numbers
of bonds, monomer-monomer interactions and bond-bond interactions in
the central square of the tree, which are given by
\begin{subequations}
\label{densi}
\begin{eqnarray}
\rho_b&=&4xf^2(1+2\omega h+2xc^\sigma+ \nonumber \\
&&\omega^3h^2+\kappa x \omega^2f^2+2x\omega^2h c^\sigma+ \nonumber \\
&&3x^2\omega\kappa c^{2\sigma})/d, \label{e10}\\
\rho_{mm}&=&4\omega(h^2+2\omega h^3+2xf^2h+ \nonumber \\
&&\omega^3h^4+3x\omega^2f^2h^2+\kappa x^2\omega f^4+ \nonumber \\
&&2x^2\omega f^2hc^\sigma+x^3\kappa f^2c^{2\sigma})/d, \label{e11}\\ 
\rho_{bb}&=&2x^2\omega\kappa f^2(\omega^2f^2+2xc^{2\sigma})/d;
\label{e12}
\end{eqnarray}
\end{subequations}
where
\begin{eqnarray}
d&=&1+4h+2h^2+4\omega h^2+4xf^2+4\omega h^3+ \nonumber \\
&&8x\omega f^2h+4x^2c^\sigma
2^2+\omega^4h^4+4x\omega^3f^2h^2+ \nonumber \\
&&2\kappa x^2\omega^2f^4+4x^2\omega^2c^\sigma
f^2h+4x^3\omega\kappa c^{2\sigma}f^2, \label{e13}
\end{eqnarray}
calculated at the fixed point of the recursion relations (Eqs. \ref{rrr}).

The thermodynamic behavior of the model is determined by the fixed
points of the recursion relations  Eqs. \ref{rrr}, each of which
correspond to a thermodynamic phase. We investigated the stability
regions for each
of the fixed points we found, which are three, in general: (a)
$a=b=0$, $c=1$, the non-polymerized fixed point, corresponding to
$\rho_b=\rho_{mm}=\rho_{bb}=0$, (b) $a, b, c \neq 0$ and finite, which is
the regular polymerized fixed point, where nonzero densities are
found, and (c)$a \rightarrow \infty$, and $b,c \neq 0$ and finite, which we 
call saturated polymerized phase, since $\rho_b=\rho_{mm}=2,
\rho_{bb}=1$ in this phase. This latter is stable in a region of the
phase diagram only for the four-coordinated Husimi lattice
$\sigma=1$, being absent in the phase diagrams of higher coordinated
Husimi lattices. To study the stability region of the saturated phase 
it is convenient to rewrite the recursion relations \ref{rrr} in terms of the
following new variables:
\begin{subequations}
\label{dv}
\begin{eqnarray}
\alpha=\frac{g_0}{g_1}, \\
\beta=\frac{g_2}{g_1},\; \\
\mbox{and } \gamma=\frac{g_3}{g_1}.
\end{eqnarray}
\end{subequations}
In these variables, the saturated fixed point is located at the origin
($\alpha=\beta=\gamma=0$). 

The stability regions of phases (a) and (c) may be 
found analytically, due to their simplicity. At the stability limit of the
non-polymerized phase the largest eigenvalue of the jacobian associated to the
recursion relations \ref{rrr} calculated at the fixed point ($a,b=0$ and
$c=1$) is equal to unity. The result is:
\begin{equation}
\kappa \leq \frac{1-2x\sigma-2x^2\sigma}{2x^3\sigma\omega}.
\label{srnp}
\end{equation}
while the stability region of the dense polymerized fixed point  is obtained
in a similar way using the recursion relations written in the other set of
variables \ref{dv}. One obtains ($\sigma=1$):
\begin{equation}
x \geq
\frac{-1+8\kappa+\omega-8\kappa\omega+4\kappa^2\omega}{\kappa\omega^3
(1-8\kappa+4\kappa^2)}.   
\label{srd}
\end{equation}
It should be remarked that for $\kappa=1+\sqrt{3/4} \approx 1.86$ the second
member of the inequality above diverges, so that the dense phase is never
stable for $\kappa \leq 1+\sqrt{3/4}$. In particular, this is true for the
model with interactions between monomers only ($\kappa=1$), as remarked by
Pretti \cite{p02}.
The stability limit of the regular polymerized phase (b) may be found
numerically.

The critical surfaces in the phase diagrams will eventually end at tricritical
lines. These lines may be obtained requiring the corresponding solution to be
a double root of the fixed point equations. Some algebra furnishes the
tricritical condition for the (a)-(b) critical surface:
\begin{eqnarray}
\label{tcnpp}
P(\omega,x)&=&-\omega + 7x - 2\omega x - 16{x^2} + 
10\omega{x^2} - \nonumber \\
&& 4(\omega-2)\omega\,{x^3} +
8(1 +(\omega-1)\omega){x^4}+ \nonumber \\
&&2(1 + 2\omega({\omega} - 1)){x^5}=0.
\end{eqnarray}
A similar calculation for the tricritical condition on the critical surface
where phases (b) and (c) are equal leads do:
\begin{eqnarray}
\label{tcpd}
{\cal P}(\omega,\kappa)&=&-2(\omega - 1)^2+
(-1+\omega)^2( 63 + 2\omega +\omega^2)\kappa- \nonumber \\  
&&8( 99 - 188\omega + 85\omega^2+ 
4\omega^4)\kappa^2+ \nonumber  \\ 
&&16( 311 - 553\omega + 217\omega^2 + 
25\omega^4)\kappa^3- \nonumber  \\
&&32( 498 - 786\omega + 213\omega^2 + 
76\omega^4)\kappa^4 + \nonumber   \\ 
&&16( 1483 - 1876\omega - 44\omega^2 + 
454\omega^4)\kappa^5 - \nonumber  \\
&&128( 112 - 120\omega - 65\omega^2 + 
76\omega^4)\kappa^6 +\nonumber  \\ 
&&128( 9 - 22 \omega - 42\omega^2 + 
50\omega^4)\kappa^7 -\nonumber \\ 
&&1024\omega^2( -1 + 2\omega^2) \kappa^8+  \nonumber \\
&& 256\omega^4\kappa^9 = 0.
\end{eqnarray}

To obtain the phase diagrams of the model, in the parameter space defined by
$x$, $\omega$, and $\kappa$, we find which fixed point is stable at each point
of the parameter space. Surfaces of this space where the stability limits of
two phases are coincident are critical surfaces, and regions where more than
one fixed point is stable are related to first order transitions. To obtain
the location of these transitions one may use a Maxwell construction, although
it is sometimes possible to find the free energy of the model on a treelike
lattice using appropriate recursion relations \cite{g95}. Due to
the simplicity of the non-polymerized and saturated phases, the partition
function per elementary square of the lattice may easily be calculated, and
this result simplifies considerably the determination of the coexistence
surface of these two phases. Since in the non-polymerized phase only
configuration 1 of figure \ref{f3} is present in the core of the lattice, the
partition function per
elementary square will be $y_{np}=1$. In the saturated phase, configuration 
11 will dominate, and therefore for this phase we have $z_s=2\kappa
x^2\omega^2$. Thus, for the four-coordinated lattice the coexistence surface,
in the region where phases (a) and (c) are stable, is given by:
\begin{equation}
\label{dnp}
2\kappa x^2\omega^2=1.
\end{equation}
Since the phases (a) and (c) have different densities, the transition
between them is always of first order, as may be verified in the phase
diagrams below.

\section{Phase diagrams}
\label{pd}
As mentioned before, for $\sigma \geq 2$ only the
non-polymerized and the regular polymerized fixed points of the
recursion relations (Eqs. \ref{rrr}) are stable, and therefore no
dense polymerized phase appears in the phase diagram. The 
non polymerized phase is stable for small values of the activity 
$x$, and as the activity is increased, eventually the regular 
polymerized phase becomes stable. The stability limits of both
phases are coincident at low values of the interaction ($\omega$ 
and $\kappa$ close to one), and thus the phase 
transition between them is continuous. As the strength of the
interaction is increased, however, the transition will become 
discontinuous, and thus the critical surface (where the both 
phases have the same densities) is separated by a tricritical 
line from the surface where both phases coexist. This phase 
diagram, where the collapse transition corresponds to a 
tricritical point, is expected for this problem since the 
theta point was recognized as a tricritical point in the 
pioneering work by de Gennes (\cite{dg75}), and thus we will 
concentrate now on the case of the four coordinated lattice.

For $\sigma=1$ the dense phase is stable in part of the 
phase diagram, and therefore richer phase diagrams are 
obtained. We will consider here constant $\omega$ cuts of the phase
diagram. Perhaps the most interesting constant $\kappa$ diagram is the one
with interactions between monomers only ($\kappa=1$), which may be found in
figure 2 in the comment by Pretti \cite{p02}. Essentially three different
types of phase diagrams are found in the $(x,\kappa)$ plane for increasing
values of $\omega$.

\begin{figure}
\includegraphics[scale=0.4]{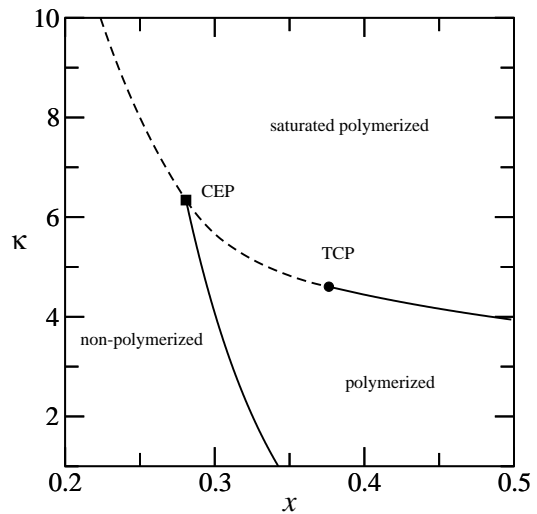}
\caption{Phase diagrams of the model for $\sigma=1$ and $\omega=1$ 
Continuous lines are second order transitions and dashed lines are first
order transitions. Tricritical points are indicated by circles and
critical endpoints are represented by squares.}
\label{f4}
\end{figure}

\begin{enumerate}
\item For $1\leq\omega\leq\omega_1$ the critical polymerization line ends at
a critical endpoint located at the confluence of the coexistence lines of the
dense phase with the other two. The particular case with interactions between
bonds only $(\omega=1)$ is depicted in figure \ref{f4}.The dense phase and the
regular polymerized phase are separated by a transition line which may be of
first or second order, a tricritical point separating these two cases. The
critical endpoint becomes a tricritical point at $\omega_1\,\simeq\, 1.15301,
\;\kappa_1\,\simeq\, 4.46985, \;x_1\,\simeq\,0.29007$. These values may
be obtained noting that this point in the parameter space is located on the
coexistence surface of the non-polymerized and the dense phases (equation
\ref{dnp}), and on the tricritical line, defined by the stability limit of the
non-polymerized phase (equation \ref{srnp} as an equality) and the tricritical
condition (equation \ref{tcnpp}).

\item For $\omega_1 < \omega < \omega_2$ tricritical points are present on the
boundaries of the polymerized phase with both the non-polymerized and the
dense phases. Three coexistence lines meet at a triple point, as may be seen
in figure \ref{f5}. As $\omega$ is increased, the tricritical point on the
boundary between the polymerized and the dense phases moves closer to the
triple point. These two points meet, becoming a critical endpoint, at
$\omega_2\,\simeq\, 1.21717, \;\kappa_2\,\simeq\, 4.08985,
\;x_2\,\simeq\,  0.28726$.

\begin{figure}
\includegraphics[scale=0.4]{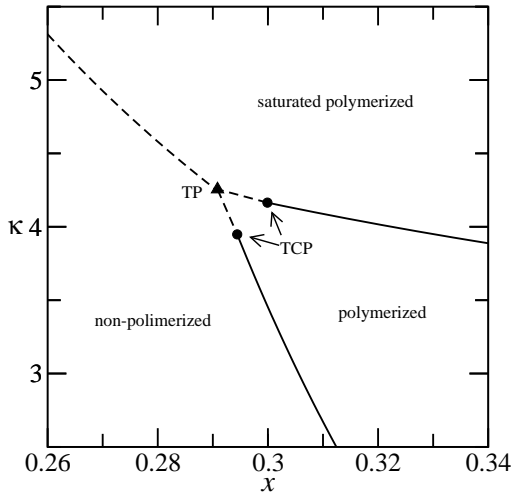}
\caption{Phase diagrams of the model for $\sigma=1$ and $\omega=1.18$. 
The triple point is indicated by a triangle.}
\label{f5}
\end{figure}

\item For $\omega \geq \omega_2$ the transition between the polymerized and
dense phases is always continuous, and this critical line ends at a critical
endpoint. An example is shown in figure \ref{f6}. The boundary between the
non-polymerized and the polymerized phases displays a tricritical point if
$\omega<1.54508$ and is always of first order for higher values of $\omega$.

\begin{figure}
\includegraphics[scale=0.4]{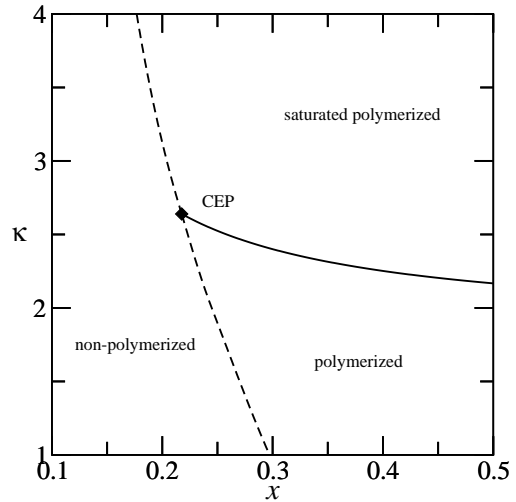}
\caption{Phase diagrams of the model for $\sigma=1$ and $\omega=2$. 
Since $\omega>1.54508$ the transition between the non-polymerized and the
polymerized phase is always of first order.}
\label{f6}
\end{figure}

\end{enumerate}
For the determination of the first order boundaries involving the regular
polymerized phase, a Maxwell construction was done using 
the pair of conjugated variables $\kappa$ and $\rho_{bb}$. Other options are
possible, this one was chosen for simplicity. The result of this calculation
is not expected to depend on the choice of the variables. For a similar model
the Maxwell relations were tested explicitly \cite{sw87} and also the free
energy was obtained directly using a proper iterative procedure \cite{g95}.

\section{Conclusion}
The solution of models for polymers with attractive interactions on
Husimi lattices built with squares leads to the expected phase
diagrams when the ramification of the lattice $\sigma$ is equal or
larger to 2. In this case, only one polymerized phase is found,
separated from the non-polymerized phase by a first- or second order
transition line. The two lines are separated by a tricritical point,
which is associated to the collapse transition of the polymers. At a
four-coordinated Husimi lattice ($\sigma=1$), however, a second
polymerized phase is stable at high values of the Boltzmann factor
$\kappa$ of the interactions between bonds. Since in this phase all sites of
the lattice are incorporated into the polymers, we called it dense
polymerized phase. 

The phase diagrams obtained for the model on the four coordinated
Husimi lattice may offer an explanation for apparently conflicting
results in the literature related to the collapse transition of
polymers on the square lattice. Transfer matrix and finite size
scaling calculations of a model with interactions between {\em
monomers} on first neighbor sites \cite{ds85} lead to compelling
evidences that, as $\omega$ is increased, the critical polymerization
line ends at a tricritical point, but exact Bethe ansatz arguments
for a magnetic model which is equivalent to a polymer model with
attractive interactions between {\em bonds} do not indicate a
tricritical collapse transition point \cite{bnw89}. In the Husimi
lattice solution of the model presented here, the collapse transition
point is a tricritical point for the case of interacting {\em
monomers} and a critical endpoint when the interaction is between
{\em bonds}. This is consistent with the results known for these
models on the square lattice. It is possible to obtain many features
of the problem of {\em directed} polymers with interacting bonds on
the square lattice exactly \cite{bovy90}. Although the collapse
transition for this case is also a multicritical point, the details
of the phase diagram are quite different from the ones found here.  A
rather unphysical feature of this model is that a polymerized phase
of zero density of monomers occupies a finite region of the phase
diagram, even in the absence of any attractive interactions.

It is of some interest to find out how the qualitatively different
phase diagrams for interactions between monomers and bonds change
into each other. Thus, we considered here a more general model where both
interactions are present. We then found
that the transition between phase diagrams were the collapse
transition is a tricritical point (interactions between monomers) and those
were it is a critical 
endpoint  (interaction between bonds) occurs through an intermediate phase
diagram were two tricritical points are present and three coexistence lines
meet at a triple point. It should be stressed that the two tricritical points
found in the phase diagram are not part of the same tricritical line in the
full parameter space $(x,\omega,\kappa)$. This poses the question if those two
distinct tricritical points actually exist for the model defined on
two-dimensional lattices and, if this is true, if they share the same set of
tricritical exponents.  
Although experimental studies of the collapse transition of polymers
confined in a two-dimensional surface have been done \cite{vr80} and
seem to confirm the tricritical nature of the transition,
monodisperse polymer solutions were used for them, while in our
calculations the chains are polydisperse.

\acknowledgments
We acknowledge Prof. Paulo Murilo C. de Oliveira, for suggesting
the study of the model in the extended parameter space. We are grateful to the
argentinian agencies CONICET and SECYTUNC, as well as the brazilian agencies
CAPES, CNPq, and FAPERJ for partial financial support. PS acknowledges the
hospitality of Universidade Federal Fluminense, where part of this work was
done.

\end{document}